\documentclass[11pt]{article}
\usepackage{amsmath}
\usepackage{amsfonts,amssymb,latexsym}

\begin{document}

\newcommand{\nl}{\nonumber\\}
\newcommand{\nnl}{\nl[6mm]}
\newcommand{\nle}{\nl[-2.5mm]\\[-2.5mm]}
\newcommand{\nlb}[1]{\nl[-2.0mm]\label{#1}\\[-2.0mm]}

\renewcommand{\leq}{\leqslant}
\renewcommand{\geq}{\geqslant}

\newcommand{\be}{\bes}
\newcommand{\ee}{\ees}
\newcommand{\bes}{\begin{eqnarray}}
\newcommand{\ees}{\end{eqnarray}}
\newcommand{\eens}{\nonumber\end{eqnarray}}

\renewcommand{\/}{\over}
\renewcommand{\d}{\partial}

\newcommand{\al}{\alpha}
\newcommand{\dlt}{\delta}
\newcommand{\eps}{\epsilon}
\newcommand{\w}{\omega}

\newcommand{\ii}{{(i)}}

\newcommand{\EE}{{\mathcal E}}
\newcommand{\N}{{\mathcal N}}

\newcommand{\hphi}{\hat\phi}
\newcommand{\hpi}{\hat\pi}
\newcommand{\hphis}{\hat\phi^*}
\newcommand{\hEE}{\hat{\mathcal E}}

\newcommand{\e}{{\mathrm e}}
\newcommand{\rank}{{\mathrm{rank}\,}}
\newcommand{\im}{{\mathrm{im}\,}}
\newcommand{\Gr}{{\mathrm Gr\,}}

\newcommand{\RR}{{\mathbb R}}
\newcommand{\CC}{{\mathbb C}}
\newcommand{\ZZ}{{\mathbb Z}}
\newcommand{\kk}{{\mathbf k}}

\title{{A BV subtlety}}

\author{T. A. Larsson \\
Vanadisv\"agen 29, S-113 23 Stockholm, Sweden\\
email: thomas.larsson@hdd.se}

\maketitle
\begin{abstract}
The standard BV complex is never acyclic provided that the equations of 
motion have solutions and the admissible class of functions is general
enough, unless one introduces second-order antifields. 
This phenomenon is explicitly illustrated for the harmonic oscillator
and the free electromagnetic field.
\end{abstract}

\bigskip

The purpose of this note is to point out a subtlety in the
treatment of the Batalin-Vilkovisky (BV) complex in the book by
Henneaux and Teitelboim	\cite{HT92}.

Recall from section 17.2.1 of \cite{HT92} how the Koszul-Tate (KT)
resolution is  constructed. For each field equation 
$\EE_i \equiv \dlt S/\dlt\phi^i = 0$, i.e. for each field $\phi^i$
(assumed bosonic for simplicity), we introduce a fermionic antifield
$\phi^*_i$. There is a grading by antifield number, defined by
$\deg \phi^i = 0$, $\deg \phi^*_i = 1$. Define the KT 
differential $\dlt$, $\deg \dlt = -1$, by
\bes
\dlt \phi^i &=& 0, 
\nlb{KT}
\dlt \phi^*_i &=& \EE_i.
\eens
The zeroth cohomology group equals the desired space of 
smooth functions over the stationary surface $\Sigma$:
\be
H^0(\dlt) = C^\infty(\Sigma) = {C^\infty(I)\/\N},
\ee
where $I$ is the space of all histories and $\N$ is the ideal generated
by the field equations $\EE_i = 0$. In the presence of gauge symmetries,
we must add extra antifields to make the KT complex acyclic. 
Each identity of the form
\be
R^i_\al \EE_i \equiv 0
\label{RE}
\ee
leads to unwanted cohomology because the linear combinations 
$R^i_\al \phi^*_i$ become closed and thus contribute to cohomology. To
kill this contribution, we must introduce further antifields 
$\theta_\al$, on which the differential acts as
\be
\dlt \theta_\al = R^i_\al \phi^*_i.
\label{dtheta}
\ee
The identities (\ref{RE}) may have further dependencies, which 
require further antifields. 

In a gauge theory, we are actually not interested in arbitrary functions
over the stationary surface, but only in the gauge-invariant ones; we
should hence factor out gauge orbits. To construct the full BV
complex, we also introduce fermionic ghosts $c^\al$, with 
$\deg c^\al = -1$, and replace the first equation in (\ref{KT}) by
\be
\dlt \phi^i = R^i_\al c^\al + more.
\ee
The differential $\dlt$ also acts on the ghosts, but we do not need
explicit formulas here.

Alas, the construction above has a subtle flaw. If we apply the
recipe (\ref{KT}) to the harmonic oscillator, there are identities
of the form (\ref{RE}) which require further antifields. So in this
sense, it appears that the harmonic oscillator has a gauge symmetry.
In fact, it is a general feature that whenever the equations of 
motion possess solutions, there are identities which require 
second-order antifields in order to make the KT complex acyclic. This
follows by a simple counting argument, given below.
In the present note we explain why and the difference between the
second-order antifields needed for solutions and gauge symmetries,
respectively.

Consider a harmonic oscillator with frequency $\w$. A history in $I$ is
a function $\phi(t)$, $t \in \RR$. The Euler-Lagrange equation reads
\be
\EE(t) \equiv \ddot \phi(t) + \w^2 \phi(t) = 0.
\ee
We therefore introduce antifields $\phi^*(t)$ and define the differential 
by
\bes
\dlt \phi(t) &=& 0, 
\nle
\dlt \phi^*(t) &=& \EE(t).
\eens
We verify the identities $\hEE(\w) \equiv 0$, $\hEE(-\w) \equiv 0$, where
\be
\hEE(k) = \int dt\ \e^{ikt} \EE(t).
\label{hE}
\ee
These identities are of the form (\ref{RE}) and thus require second-order
antifields.

The calculations become even clearer if we pass to Fourier space. A 
history in $I$ is now described by a function $\hphi(k)$, $k \in \RR$,
and the equations of motion take the form
\be
\hEE(k) \equiv (k^2 - \w^2)\hphi(k) = 0.
\label{sol}
\ee
The stationary surface $\Sigma$ is spanned by $\hphi(\w)$ and 
$\hphi(-\w)$, so we can identify $C^\infty(\Sigma)$ with the space
of smooth functions of the form $F(\hphi(\w), \hphi(-\w))$.
The differential acts as
\bes
\dlt \hphi(k) &=& 0, 
\nlb{KTh}
\dlt \hphis(k) &=& (k^2 - \w^2)\hphi(k).
\eens
The advantage of the Fourier transformation is that the different modes 
decouple, and we may consider 
each $k$ separately. For $k^2 \neq \w^2$, $\hphi(k)$ is closed but also
exact, since $\hphi(k) = \dlt(\hphis(k)/(k^2 -\w^2))$, whereas 
$\hphis(k)$ is not closed. Hence both $\hphi(k)$ and $\hphis(k)$
vanish in cohomology. In constrast, if we specialize (\ref{KTh}) to
$k^2 = \w^2$, say $k = +\w$, we have
\be
\dlt \hphi(\w) = \dlt \hphis(\w) = 0.
\ee
Hence both $\hphi(\pm\w)$ and $\hphis(\pm\w)$ are closed but not exact,
and both contribute to cohomology. Using the Grassmann nature of the 
antifields, it is clear that the only nonzero cohomology groups are
\bes
H^0(\dlt) &=& \{ F(\hphi(\w), \hphi(-\w)) \} \equiv C^\infty(\Sigma), \nl
H^1(\dlt) &=&  \{ F_1 \hphi^*(\w) + F_2 \hphi^*(-\w) :
F_1, F_2 \in C^\infty(\Sigma) \}, 
\label{HHH}\\
H^2(\dlt) &=&  \{ G \hphi^*(\w) \hphi^*(-\w) : G \in C^\infty(\Sigma) \}.
\eens
The unwanted cohomology groups originate from the identities
$\hEE(\w) = \hEE(-\w) = 0$. To kill them, we follow the recipe in 
(\ref{RE}) and (\ref{dtheta}) and introduce two second-order 
antifields $\theta(\w)$ and $\theta(-\w)$, on which $\dlt$ act as
\bes
\dlt \theta(\w) &=& \hphi^*(\w), 
\nlb{dthetaw}
\dlt \theta(-\w) &=& \hphi^*(-\w).
\eens
This modified definition of $\dlt$ yields a resolution of 
$C^\infty(\Sigma)$: $H^0(\dlt) = C^\infty(\Sigma)$ and $H^n(\dlt) = 0$
for all $n \neq 0$.

It has been pointed out by U. Schreiber that the unwanted
cohomology in (\ref{HHH}) does not exist if the space of smooth
functionals $C^\infty(I)$ is defined in a sufficiently restrictive
manner. If the space of histories $I$ were finite-dimensional or
discretely infinite-dimensional, there is only one reasonable definition
of smoothness: $F \in C^\infty(I)$ if $F(\hphi)$
is a smooth function of its argument. However, $I$ is continuously
infinite-dimensional, and one may additionally require that every
functional $F[\hphi(k)]$ depends smoothly on $k$. This extra
condition does not affect $H^0(\dlt) = \ker\dlt_0/\im\dlt_1$, which
is a quotient of spaces of functionals which are smooth in this
stronger sense, but it makes the higher cohomology groups vanish;
e.g., $\hphi(\w) \propto \hphi^*(k)\dlt(k-\w) \in \ker\dlt_1$ is not
a smooth function of $k$.
Nevertheless, such a narrow definition of smoothness is quite
unnatural. In the time domain, it corresponds to ``adiabatic''
functionals that are independent of $\phi(t)$ at
$t = \pm\infty$; this condition rules out $\hEE(k)$ defined as in
(\ref{hE}). Moreover, it is irrelevant for compactified time, where 
continuity in $k$ is no longer an issue. Therefore, it is more useful to 
define $C^\infty(I)$ to include delta-function distributions in $k$. 
The relevant associative 
product is the convolution product, corresponding to pointwise
multiplication in the time domain. Throughout this paper, we
consider a function space in which delta-functions in $k$ are well
defined. The second-order antifields in (\ref{dthetaw}) may be viewed
as the necessary correction when we relax from the stronger
notion of smoothness.
 
To kill a gauge symmetry, one would not only introduce a second-order
antifield but also a fermonic ghost with antifield number $-1$; the 
purpose of this ghost is to identify points on gauge orbits. We could
consistently do this for the identities (\ref{sol}) as well; introduce
two fermionic ghosts $c(\w)$ and $c(-\w)$, with
$\dlt c(\w) = \dlt c(-\w) = 0$, and replace the first equation in
(\ref{KTh}) by
\be
\dlt \hphi(k) = c(\w) \dlt(k-\w) + c(-\w) \dlt(k+\w).
\ee
Although consistent, this modification is not desirable, since it makes
$H^\bullet(\dlt)$ vanish completely.

The existence of identities of the type (\ref{sol}) is quite general and
not particular to the harmonic oscillator; it is a conseqence of the
equations of motion having solutions. This can be seen by a simple
counting argument, most clearly formulated in a finite-dimensional
context. Hence we replace the base space $\RR$ by a finite lattice with
periodic boundary conditions; differential equations turn into 
difference equations. 

Consider a lattice with $n$ points,
and assume that the equations of motion are linear for simplicity.
The field is now an $n$-dimensional vector $u$, subject to a matrix 
equation $Au = 0$, where $A$ is an $n\times n$ matrix of rank $n-p$.
Hence the equations of motion have $p$ independent solutions. Now 
introduce an $n$-dimensional antifield vector $u^*$, and define the 
differential by
\bes
\dlt u &=& 0, 
\nlb{lattice}
\dlt u^* &=& Au.
\eens
Since $\rank A = n-p$, there is an $(n-p)$-dimensional subspace where
$A$ can be inverted; on this subspace, $u = \dlt(A^{-1} u^*)$ is exact.
But this also means that vectors of the form $Au$ only span an
$(n-p)$-dimensional subspace, and every $u^*$ such that $\dlt u^*$ is 
perpendicular to this subspace is closed. Consequently, $H^\bullet(\dlt)$
is generated by $p$ $u$'s and $p$ $u^*$'s.

Define the Grassmann number $\Gr$ as the
difference between the number of bosonic and fermionic degrees of 
freedom, i.e. we set $\Gr u = 1$ and $\Gr u^* = -1$. The space of 
histories $I$ is spanned by $n$ $u$'s and $n$ $u^*$'s, and hence
the total Grassmann number is zero. Cohomology kills fields in pairs: for
each $u$ which is not exact, there is a $u^*$ that is not closed, and
vice versa. Hence passage to cohomology preserves the Grassmann number,
which must remain zero. Indeed, $H^\bullet(\dlt)$ is generated by $p$
$u$'s and $p$ $u^*$'s. To kill the latter in cohomology, we need to
introduce $p$ bosonic second-order antifields $\theta$ with 
$\Gr \theta = +1$. Then the total Grassmann number equals $p$, both 
before and after passage to cohomology.

This counting argument is completely general, and works for
finite-dimensional vectors and fields over spacetime alike, provided that
we work in a sufficiently general space of functions, as discussed above.
It also works for nonlinear equations of motion. To end up with a
cohomology with nonzero Grassmann number $p$, we must start with
Grassmann number $p$. Since the fields and antifields in (\ref{KT}) or in
(\ref{KTh}) together have Grassmann number zero, some antifields must
survive in cohomology if some fields do. We can only kill this unwanted
cohomology by adding bosonic antifields by hand.

The difference between the identities due to solutions and to genuine 
gauge symmetries can be illustrated by electromagnetism in four
dimensions. In Fourier space, the relevant field is the gauge potential
$A_\mu(k)$, $k \in \RR^4$, $\mu = 0,1,2,3$. The equations of motion,
\be
\hEE_\mu(k) \equiv k^2 A_\mu(k) - k_\mu k^\nu A_\nu(k) = 0,
\ee
are subject to two classes of identities. On the one hand we have those
due to genuine gauge symmetries, 
\be
k^\mu \hEE_\mu(k) \equiv 0,
\label{kE}
\ee
for all $k \in \RR^4$.
But we also have those due to solutions. For each lightlike vector $k$,
introduce vectors $\eps_\ii^\mu(k)$ perpendicular to it. These vectors
are thus assumed to satisfy
\be
k^2 = 0, \qquad k_\mu \eps_\ii^\mu(k) = 0.
\label{eps}
\ee
It is clear that
\be
\eps_\ii^\mu(k) \hEE_\mu(k) \equiv 0,
\label{epsE}
\ee
for each lightlike $k$. There are three vectors $\eps_\ii^\mu(k)$
satisfying the conditions (\ref{eps}), but one of them is proportional
to $k$ itself, and the corresponding identity (\ref{epsE}) is already
taken care of by the gauge identity (\ref{kE}). Therefore the index
$i$ runs over the two transverse directions $i=1,2$. In particular, for
a photon moving along the $z$ axis, $k = (1,0,0,1)$, 
$\eps_{(1)}(k) = (0,1,0,0)$ and $\eps_{(2)}(k) = (0,0,1,0)$.

Hence we must introduce two kinds of antifields to cancel the two types of
spurious cohomology generated by (\ref{kE}) and (\ref{epsE}). This example
illustrates the crucial difference. The gauge identity (\ref{kE}) holds
for all $k \in \RR^4$, whereas the solution identity (\ref{epsE}) only
holds if $k^2 = 0$; the space of such $k = (k_0, \kk)$ is labelled by 
$\kk \in \RR^3$. Putting it all together, we introduce the
bosons $A_\mu(k)$, $\zeta(k)$ and the fermions $c(k)$, $A^*_\mu(k)$, 
defined for all $k \in \RR^4$, and additional bosons 
$\theta_\ii(k)$, $i = 1,2$, only defined for $k^2 = 0$. The BV 
differential acts as
\bes
\dlt c(k) &=& 0, \nl
\dlt A_\mu(k) &=& k_\mu c(k), \nl
\dlt A^*_\mu(k) &=& k^2 A_\mu(k) - k_\mu k^\nu A_\nu(k), \\
\dlt \zeta(k) &=& k^\mu A^*_\mu(k), \nl
\dlt \theta_\ii(k) &=& \eps^\mu_\ii(k) A^*_\mu(k),
\qquad \hbox{$i = 1, 2$ and $k^2 = 0$}.
\eens
The second-order antifields $\theta_\ii(k)$ clearly provide the correct
surplus of bosonic degrees of freedom.

As another example, consider a massless free field in two dimensions.
The equation of motion reads $\d_z \d_{{\bar z}} \phi(z,\bar z) = 0$.
Define the Taylor coefficients $\phi_{mn}$ by
\be
\phi(z,\bar z) = \sum_{m,n=0}^\infty \phi_{mn} z^m \bar z^n.
\ee
The KT complex takes the form
\bes
\dlt \phi_{mn} &=& 0, \nl
\dlt \phi_{mn}^* &=& mn \phi_{mn}, \nl
\dlt \theta_m &=& \phi_{m0}^*, \\
\dlt \bar\theta_n &=& \phi_{0n}^*, \nl
\dlt \chi &=& \theta_0 - \bar\theta_0.
\eens
Note the presence of second-order antifields $\theta_m$ and 
$\bar \theta_n$, and the third-order antifield $\chi$.
$H^0(\dlt)$ is generated by $\phi_{m0}$ and $\phi_{0n}$, and the
other cohomology groups are empty, as they should. In particular,
$\chi$ is necessary to avoid double counting of $\phi_{00}$. In 
this  example, the set of Taylor coefficients is discrete, and 
no problems with delta-functions arise.

It seems surprising that the existence of spurios cohomology for the BV
complex should not have been noticed by other authors, since it occurs
already for the harmonic oscillator. However, I have never seen this
issue discussed elsewhere, so to the best of my knowledge this is
unknown, or at least not widely known. Note also that the flaw in the
usual treatment is in some sense small; BV cohomology can be regarded as
a trick to construct $H^0(\dlt) = C^\infty(\Sigma)$, which does come out
right even without extra antifields. It is nevertheless a nuisance that
$H^n(\dlt) \neq 0$ for $n \neq 0$, and it can be easily fixed by adding
some extra antifields, as illustrated for the harmonic oscillator,
electromagnetism, and the massless scalar field in two dimensions.

The problem with extra cohomology was first noted in \cite{Lar04}. The
goal in that paper was to adapt the BV formalism to canonical
quantization; the strategy was to quantize in the history phase space
first and apply dynamics as a constraint \`a la BRST afterwards. However,
to do canonical quantization, we need an honest Poisson bracket and not
just an antibracket, and therefore we must introduce momenta canonically 
conjugate to the fields and antifields; for the harmonic
oscillator these were denoted by $\hpi(k)$ and $\hpi^*(k)$, respectively. 
The problem with unwanted cohomology then becomes acute, because 
expressions like $\hphi^*(\w)\hat\pi^*(\w)$ belong to the degree zero 
subspace and hence contribute to $H^0(\dlt)$. The correct treatment, at 
least for the harmonic oscillator, appeared in \cite{Lar07}.

To conclude, we noted in this paper that extra antifields must be
introduced to make the BV complex acyclic even in the absense of gauge
symmetries, because otherwise unwanted cohomology is generated at nonzero
degree, provided that the equations of motion have solutions. This was
shown explicitly for the harmonic oscillator. In the case of a non-compact
time dimension, the extra cohomology can also be avoided by restricting 
attention to ``adiabatic'' functionals which depend smoothly on $k$, i.e.
are turned off when $t \to \pm\infty$. However, such a solution is not
very satisfactory, because it does not work if time is compact, nor in
the non-compact case if we permit non-adiabatic functionals. 
Despite the similar
treatment of gauge symmetries and solutions, there is a difference between
the corresponding antifields: a gauge symmetry depends on arbitrary
functions on spacetime, whereas a solution only depends on arbitrary
functions on a simultaneity surface.

I thank U. Schreiber for a discussion.

\end{document}